# Classifying Epistemic Relationships in Human-AI Interaction: An Exploratory Approach

Shengnan Yang[1], Rongqian Ma[2]

[1] University of Western Ontario, Canada
[2] Indiana University Bloomington, USA

**Abstract**

*As AI systems become integral to knowledge-intensive work, questions arise not only about their functionality but also their epistemic roles in human–AI interaction. While HCI research has proposed various AI role typologies, it often overlooks how AI reshapes users' roles as knowledge contributors. This study examines how users form epistemic relationships with AI—how they assess, trust, and collaborate with it in research and teaching contexts. Based on 31 interviews with academics across disciplines, we developed a five-part codebook and identified five relationship types: Instrumental Reliance, Contingent Delegation, Co-agency Collaboration, Authority Displacement, and Epistemic Abstention. These reflect variations in trust, assessment modes, tasks, and human epistemic status. Our findings show that epistemic roles are dynamic and context dependent. We argue for shifting beyond static metaphors of AI toward a more nuanced framework that captures how humans and AI co-construct knowledge, enriching HCI's understanding of the relational and normative dimensions of AI use.*

**Keywords:** Epistemic Relationship, Human-AI Interaction (HAI) and Collaboration, AI Metaphor, Semi-structured Interview

## 1. Introduction

Artificial intelligence (AI) systems have made remarkable advances in recent years, often outperforming humans in complex cognitive tasks such as reasoning, diagnosis, and decision-making. These developments have sparked debate over their implications for human agency, often framed as a tension between technological determinism and human-centered views. Some warn that over-reliance on AI may erode human epistemic agency—the capacity to actively participate in knowledge creation. Others, such as Felin and Holweg (2024), emphasize the fundamental difference between AI's data-driven prediction and human theorizing. Meanwhile, practice-oriented scholars in HCI and management highlight the benefits of human–AI interaction, framing AI as a tool for automating routine work or augmenting human decision-making (e.g. Heer 2019; Taudien et al., 2022). Research on hybrid agency and human–AI ensembles has shown that combining human and algorithmic judgments can enhance performance in certain contexts, such as transportation (Kahr et al, 2025), management (Csaszar et al., 2024), and health (Choudhary et al., 2025). However, these discussions remain largely focused on functional capabilities, cognitive mechanism and task outcomes, paying limited attention to the epistemic dimension of human–AI interactions (HAI).

Yet across these perspectives, the epistemic framing of human–AI interaction and collaboration remains underdeveloped, particularly in HCI research. As AI systems are increasingly embedded in knowledge-intensive domains, ranging from information retrieval and content generation to scientific modeling and decision support, little attention has been paid to how humans and AI co-produce knowledge or how AI reshapes users' roles as knowers. In contrast, recent philosophical scholarship has extended concerns about epistemic injustice and asymmetry, traditionally explored in human-to-human contexts, into the domain of human–nonhuman relations. AI is increasingly theorized not merely as a cognitive tool supporting mental processes, but as an epistemic technology that contributes to knowledge acquisition and, under certain conditions, may even function as an epistemic authority (Ferrario et al., 2024; Hauswald, 2025). These accounts challenge traditional human-centered assumptions and propose an alternative lens for conceptualizing human–AI interaction as a form of epistemic co-agency or negotiated collaboration. Such views raise new questions about whether epistemic relationships can form between humans and AI, even when AI lacks beliefs, intentions, or moral agency (Ryan, 2020).

In this study, we define an epistemic relationship as the interactional pattern through which human users evaluate, rely on, or negotiate the epistemic contributions of AI systems in knowledge work. Rather than assuming fixed roles or static attitudes, we examine how such relationships are dynamically constructed across contexts. Drawing on 31 semi-structured interviews with academics working with AI in their work, we ask:

**RQ1:** Does an epistemic relationship exist during human-AI interaction?

**RQ2:** If so, what are the main patterns of human-AI epistemic interaction?

## 2. Literature Review

### 2.1. Framing AI Roles in HCI – Functional, Yet Epistemically Thin

HAI research has developed a wide array of typologies describing the roles AI systems play within task-based and decision-oriented workflows. Across HCI study, AI has been labeled as a tool, assistant, servant, oracle, collaborator, partner, or guide (Papachristos et al., 2021; Kim et al., 2021; Kim et al. 2024). These framings are typically grounded in technological functionality and behavioral affordance.

These functional classifications also inform adjacent concerns, mainly including user perception (Hwang & Won, 2022), cognitive factors (Divis et al., 2022; Taudien et al., 2022), performance evaluation (Fragiadakis et al., 2025), and trust calibration (Mehrotra et al., 2024; Kahr et al., 2025), often through the lens of technology adoption models (e.g., Sundar, 2020; Glikson & Woolley, 2020; Pareek et al., 2025). Constructs such as perceived agency, sense of control and trustworthiness recur across this literature, highlighting ongoing attention to automation boundaries and power dynamics in human–machine interaction (including but not limited to AI) (Cornelio et al., 2022; Hwang & Won, 2022; Limerick et al., 2014). Papachristos et al. (2021) show how users' attribution preferences and self-assessed competence shape reliance patterns, while Pareek et al. (2025) demonstrate how causal explanations affect trust calibration. Kahr et al. (2025) find that experts may withhold trust from high-performing AI to preserve human values—hinting at deeper epistemic judgments. However, these studies often remain outcome-oriented and under-theorized in epistemic terms. The concept of trust itself is frequently ill-defined or treated instrumentally, rather than examined through normative or epistemological lenses.

This orientation results in a conceptual gap. Most framings treat AI as an input–output machine positioned within fixed interface boundaries, rather than a participant in knowledge construction. Even where "agency" or "trust" is discussed, the focus is typically on perception and delegation, not on how AI systems reconfigure the user's epistemic stance—as knowers, validators, or co-constructors of knowledge. In this sense, HCI remains epistemically thin: it lacks a clear framework for analyzing how AI systems mediate or reshape knowledge relationships.

Yet, epistemic concerns are not entirely absent. Many HCI discussions implicitly echo philosophical concepts such as epistemic agency, authority, and dependence, though rarely named as such. For example, perceived AI autonomy (Kim et al., 2024) maps closely onto epistemic agency. Papachristos et al. (2021) show that users may see the same AI as a confirming mirror, helpful assistant, or trusted oracle—implying variable epistemic status. Similarly, the notion of epistemic authority is evoked in trust studies that assess AI expertise or anthropomorphic framing, yet seldom explore when such authority is justified. Dependence on AI outputs—manifested in automation bias—is widely documented but rarely analyzed in terms of its long-term epistemic implications, such as the erosion of user agency. A notable exception is Wu et al. (2024), who explicitly examine shared epistemic agency in GenAI-facilitated learning and point toward a more symbiotic epistemic relationship between humans and machines, one in which both parties' contributions are recognized and epistemically valued.

In sum, while HCI has made substantial progress in classifying AI roles and understanding user perceptions, its theoretical treatment of human–AI relations remains fragmented and predominantly functional. The implicit epistemic questions embedded in trust, agency, and attribution call for a more explicit epistemic framing—one that can account for how AI systems participate in, mediate, or challenge knowledge construction. To address this, the next section introduces an epistemic lens mainly from recent philosophical study of technology that offers us the potential to reconnect these scattered insights into a more coherent framework for analyzing human–AI knowledge relations.

### 2.2 Epistemic Perspectives on Human–AI Interactions

Although traditionally limited to human-to-human interactions, epistemic concerns are increasingly being raised in relation to non-human agents. In particular, generative AI has prompted renewed inquiry into whether such systems can hold or exercise epistemic capacities. As Alvarado (2023) notes, artificial intelligence is best understood as an epistemic technology, a view that opens new directions for theorizing its role in knowledge practices. We briefly review this philosophical turn to lay the groundwork for our own analytical framework on human–AI epistemic relationships.

**Epistemic Nature of AI: Epistemic Technology and Epistemic Authority** Recent philosophical scholarship has increasingly framed AI through an epistemic lens, emphasizing their role not merely as computational tools, but as epistemic agents embedded in knowledge-producing practices. AI, in particular, are seen as both cognitive and epistemic technologies (Alvarado 2023; Heersmink et al., 2024): cognitive, in that they assist with summarizing, classifying, or translating; epistemic domains, broader categories of practices in that they provide information, answer questions, and generate outputs that support knowledge acquisition, retention and use.

Alvarado (2023) argues that epistemic technologies can be conceptually and practically distinguished from other technologies based on what they are designed for, what they do, and how they do it. He contends that AI can be uniquely positioned as an epistemic technology because it is developed and deployed for epistemic contexts such as inquiry, applied to manipulate epistemic content, and operates on such content through processes like inference, prediction, and analysis. Building on this, Ferrario et al. (2024) argue that AI systems—especially generative models—function as epistemic technologies whose behavior is shaped not solely by designer intent, but also by training data. These systems can act as epistemic enhancers, augmenting users' ability to access and apply knowledge. Crucially, this enhancement is context-sensitive and interaction-dependent.

This raises the further question: under what conditions might AI systems be granted epistemic authority? Hauswald (2025) proposes a non-anthropocentric model of epistemic authority grounded not in intentional states or mental beliefs but in the suitability of an output as a truth-indicator. On this view, epistemic deference is justified not because an agent has beliefs, but because its outputs reliably track truth in a given domain. When users encounter outputs that are accurate, coherent, and contextually appropriate, they may justifiably defer to them much like to trusted experts. Importantly, epistemic authority is not static—it is a situational and revisable stance, contingent upon the user's task, context, and evaluative standards. This offers a more dynamic understanding of AI's place within human knowledge construction.

**Attributes of Epistemic Relationships in Human–AI Interaction.** Hauswald's model contributes in three key ways. First, it decouples epistemic authority from anthropocentric assumptions, allowing for non-human agents to occupy epistemic roles. Second, it emphasizes that such authority is not an inherent attribute but emerges through situated interactions with users. Third, and most critically, it disentangles epistemic authority from trust, enabling a sharper conceptual focus on how judgments of epistemic reliability are formed.

This move toward separating epistemic authority from trust aligns with a broader trend. A few scholars (Blanco et al., 2022; Ferrario et al., 2024) have explored more nuanced definitions of trust and its construction process. For instance, according to Blanco et al.'s (2022) relational theory of trust, trust is not grounded in interpersonal morality or anthropomorphic expectations, but rather shaped by context and task-specific epistemic positioning. In this framework, trust is not a prerequisite for epistemic authority, but a parallel lens—one through which human–AI interaction is framed and negotiated. On this account, trust and authority are distinct but co-occurring dimensions in human–AI interaction, shaped by how users frame their expectations in specific tasks and domains.

This shift also foregrounds the question of assessment: how do users determine the appropriateness or reliability of AI outputs? While outcome-based evaluation remains common, recent work (e.g., Russo et al., 2024) suggests that users also assess underlying epistemic processes, such as transparency, consistency, or alignment with disciplinary norms. This process-based assessment complicates but enriches our understanding of how epistemic relationships between humans and AI unfold.

Ferrario et al. (2024) extend this line of thought by introducing the idea of hybrid epistemic agency, where human cognition and AI outputs are entangled in collaborative knowledge practices. However, this hybridity does not erase epistemic asymmetry. On the contrary, it can produce new forms of epistemic dependence, where users defer to AI outputs without understanding the underlying logic. Here, the expertise dimension becomes especially salient. Asymmetries between experts and lay users shape the evaluation criteria applied, the likelihood of critical engagement, and the readiness to defer to AI-generated results. While the expert–layperson distinction is well-known in classical epistemology, its significance is renewed in the context of generative AI, where systems increasingly mediate knowledge production and validation.

Russo et al. (2024) further refine the distinction by identifying two forms of epistemic stance. In epistemic symmetry, expert users evaluate AI outputs by directly interrogating their generative processes—for instance, by examining their logic, fairness, or technical soundness. In epistemic asymmetry, by contrast, non-experts must rely on indirect cues or institutional guarantees. These two configurations not only shape how users assess AI but also influence the types of epistemic trust they develop.

Together, these three dimensions—assessment perspective, trust type, and human epistemic status (expert or not)—form the analytical foundation for the five epistemic relationships (ERs) we identify in the next section. They jointly shape how AI systems are positioned within knowledge practices, such as tools, assistants, collaborators, or authorities, etc.

## 3. Methods

Given this study's purpose of understanding human–AI epistemic interaction, we adopted an exploratory methodological approach. To support a more comprehensive and credible analysis, we employed data triangulation by combining and reanalyzing two independently collected datasets to generate novel insights. Both datasets were collected in 2024 and 2025 through semi-structured interviews with academics and centered around AI use and HAI in research and teaching contexts. Each interview, conducted either in person or via Zoom, lasted approximately one hour and was audio-recorded for analysis. Participants were asked about the tasks they use AI for, their perceptions of AI's role, and how they evaluate the quality and outcomes of AI involvement. The final dataset includes 31 participants from over 10 countries and regions, including the U.S., China, the U.K., Mexico, and various EU countries, and spans diverse fields such as digital humanities, data science, and information and library science.

To analyze the data, we developed a five-category codebook through an iterative process of open coding and synthesis of relevant literature in the philosophy of science, science and technology studies, and HAI. The codebook, as shown in Table 1, captures key dimensions of human–AI interaction and provides a structured framework for examining how AI is used, assessed, and understood in academic contexts.

In this codebook, "Metaphor" categorizes how humans think about AI as an entity and assesses its epistemological status—whether they perceive AI primarily as a mere tool, a supportive assistant (with lesser cognitive contributions), an active collaborator (co-agent with equal cognitive contributions), a coach or mentor (with superior cognitive contributions), or even as an epistemic authority (oracle) (Hauswald, 2025). "Task Type" identifies the specific cognitive scenarios and purposes in which AI is deployed. Following the tasks, the codes of "Assessment Perspective" and "Human Epistemic Status" provide insights into how humans evaluate the performance of AI outputs. "Assessment Perspective" indicates whether the evaluation focuses on specific outcomes generated by the AI, or the overall quality of the processes involved. In addition, "Human Epistemic Status" denotes the evaluator's position, indicating whether the human collaborator is an AI expert, a domain expert, both, or a layperson in the task.

*Trust Type* refers to how humans interpret their relationship with AI, offering one lens to understand its underlying dynamics (Blanco, 2025). Four distinct types of trust emerged from the coding process: *Distrust* means both the absence of trust and a confident negative expectation about another's conduct. *Reliance* describes task-based dependence on AI without emotional or moral expectations. It focuses solely on outcomes, not methods. For example, a researcher may rely on AI to process large datasets quickly, without questioning how the results are generated. *General trust* involves a broad, informal belief that AI will perform well, often based on reputation rather than evidence. For instance, a researcher might use AI in their work because colleagues recommended it or it's well-regarded in the literature, without exploring how it works. *Appropriate trust*, in comparison, is grounded in justified confidence in both the AI's results and its methods. A researcher with appropriate trust would evaluate the AI's methodology and confirm it meets disciplinary standards, trusting not only the outcome but also the reasoning behind it.

Using this five-category codebook, each author independently coded half of the interview transcripts. We then met to resolve ambiguous cases through collaborative review and discussions. Building on these results, we adopted a bottom-up, data-driven approach to further identify *epistemic relationships* (ERs), analyzing user narratives across three mutually exclusive dimensions: human epistemic status (e.g., domain vs. AI experts), assessment perspective (e.g., process vs. outcome-based evaluation), and trust type (e.g., distrust, reliance, appropriate, general), along with associated tasks and metaphors. The ERs are not exhaustive combinations of these dimensions, but empirically grounded patterns that emerged as stable clusters in how users negotiate their epistemic relationships with AI across various contexts.

**Table 1.** Codebook

| Code Category | Definition | Value | Theoretical Reference |
|---|---|---|---|
| **Metaphor** | Judgment about the epistemic status assigned to AI | Tool, Assistant, Co-agent, Coach/Mentor, Authority (oracle) | Kim, et al. (2023) Papachristos, et al. (2021) |
| **Task Type** | Specific task types in research where AI is utilized, defining the cognitive scenario and usage purpose. | E.g., brainstorming, literature search & review, summarization, data extraction, coding & programming, analysis, content generation, translation, writing, editing, general search, teaching prep, critique | N/A |
| **Assessment Perspective** | How humans evaluate the involvement of AI in their work process | Outcome-based (e.g., accuracy, quality of results), Process-based (e.g., algorithms), Mixed | Russo et al. (2024) |
| **Human Epistemic Status** | Identity and position of the human involved in the task, which may influence their AI role assignment and evaluation. | AI expert only, Domain expert only, Both, Non-expert (layperson) | Russo et al. (2024) |
| **Trust Type** | Humans' willingness to rely on AI, shaped by factors such as evaluation of its performance, perceived reliability, and understanding (or lack thereof) of its underlying processes. | Distrust, Reliance, Appropriate trust, General trust | Blanco (2025) |

## 4. Results

Our study finds that epistemic relationships do exist in many scenarios where AI is employed as an epistemic technology. Across a range of research and teaching contexts, we observed from the interviews that humans interact with AI systems not only as tools, but as cognitive partners involved in knowledge-related tasks such as ideation, synthesis, analysis, and interpretation. These interactions often involve users attributing epistemic roles to AI, relying on its outputs, challenging its responses, or co-constructing meaning through iterative engagement. Such activities reveal that epistemic relationships are not only present in HAI but actively shape how users assess, trust, and collaborate with AI in knowledge production.

Additionally, our analysis identified five specific ERs, which highlight patterns of AI's epistemic participation in human-AI interaction and show a dynamic and context-dependent character of epistemic relationships with AI. This dynamic character means that the relationships are not fixed but are shaped through ongoing negotiation across different tasks and goals. Users' epistemic expectations and trust in AI varied based on task type (e.g., coding, translation, idea generation), modes of assessment (e.g., outcome-based vs. process-based), and the perceived stakes involved. Based on these findings, we suggest moving away from static classifications of AI's role (metaphor) in research and toward an analytic framework that attends to the interplay between metaphors and task type, trust type, assessment perspective, as well as human epistemic status. Our work shows that this perspective captures the situated nature of epistemic relationships with AI and supports a more nuanced understanding of how these relationships are constructed and adjusted in research and pedagogical contexts.

**Table 2.** Five Epistemic Relationships of Human-AI Interaction

| Category | Name | Brief Definition | Metaphor | Trust Type | Assessment Perspective | Human Epistemic Status | Tasks Covered |
|---|---|---|---|---|---|---|---|
| ER-1 | Instrumental Reliance | Relies on AI solely for task efficiency; assigns no epistemic status to AI | Tool | Reliance | Outcome-based | Domain expert | Editing, translation, coding/ programming |
| ER-2 | Contingent Delegation | Delegates judgment to AI in specific tasks while maintaining oversight | Assistant / Co-agent | Appropriate trust | Outcome-based | Domain expert | Summarization, content generation, general search |
| ER-3 | Co-agency Collaboration | Treats AI as a collaborative thinker in knowledge construction | Co-agent / Mentor | General / Appropriate trust | Mixed (process/ outcome) | Domain or AI expert | Writing & rewriting, brainstorming, teaching prep |
| ER-4 | Authority Displacement | Grants AI partial authority in knowledge production | Co-agent / Mentor | General trust | Process-based | Domain/ AI hybrid | Critique, brainstorming |
| ER-5 | Epistemic Abstention | Uses AI but denies its epistemic contribution | Tool | Distrust | Outcome-based | Both | Summarization, General search |

**ER-1** reflects a utilitarian view of AI, in which users treat it as a tool to increase task efficiency without granting it epistemic status. Assessment is strictly outcome-based (e.g., whether the task gets done), and trust remains at the level of functional reliability, not belief. This relationship is most common among domain experts performing tasks such as programming or editing, where AI is employed to reduce workload or streamline processes. For example, one digital humanities expert uses AI to assist him with programming, which he is unfamiliar with and finds time-consuming. As he explained below, AI functions as a programming assistant, enabling him to achieve his goals more efficiently. With no background in computer science, he primarily evaluates AI's performance based on the outcomes, and his reliance on AI as a tool developed gradually through a trial-and-error approach:

> "I am almost exclusively using AI for data analysis—writing Python scripts to collect web data and build databases. I was never trained in computer science; everything from GIS to Python is self-taught. That's where generative AI is enormously helpful—it lets me do things I only need a few times a year. Because I don't code regularly, I often forget what I've learned and need to relearn it. AI helps me bridge that gap. What would've taken two weeks now takes two hours. I can quickly go from a lower to upper intermediate coder and focus on the logic of the algorithm instead of syntax or data types. It makes me more willing to take on Python-heavy projects because I know I won't get bogged down." (P03-DH)

In another case, the participant reported using AI to check and correct the grammar in her papers, which effectively improved both her writing and her confidence as a non-native English speaker (P27-DH).

Compared to ER-1, **ER-2** involves selectively delegating cognitive tasks to AI under specific conditions while retaining human supervision. In this case, AI is framed as an assistant or co-agent, and users exhibit appropriate but conditional trust. Users primarily evaluate its performance based on the outcome, not its reasoning process. The interaction typically does not require multi-turn engagement. The user monitors outputs and intervenes only to adjust or refine the outputs as needed. For instance, P08-DH described her interaction with AI as follows:

> "I'm still going back and forth with it, trying to see if it will reach a point where I can trust it. But for me, that trust only develops after I check everything it produces. It's like working with a research assistant: if you hire an undergraduate to help and they hand you a nicely polished essay, you're not just going to accept it as-is. You'll need to go through it, add comments, and make sure they did a good job."

Different from the contingent nature of ER-2, **ER-3** describes a more integrated and dynamic epistemic relationship in which users treat AI as a thinking partner or mentor. In this scenario, trust generally happens at a higher level and in a more flexible form, shaped by ongoing interaction rather than predetermined roles.

Additionally, assessment perspectives combine both process and outcome considerations. This configuration is evident in more interpretive and open-ended tasks, such as exploratory writing, translation, lesson planning, or brainstorming, where users draw on domain and/or AI expertise.

For example, P02-DH described a case in which AI was engaged as a co-agent partner by an early modern scholar working on the translation of diplomatic correspondence between Italy and the Dutch Republic. Although the scholar is an expert Latinist, when uncertain about specific translations, he would "treat GPT like a colleague" and consult it with Latin translation questions:

> "He'll say, 'What do you think about this?' or 'I think it might mean this, but I'm not sure.' Then GPT responds, saying something like, 'I suggest this translation, although this might also be possible.' He's essentially having a conversation with it. He has completely anthropomorphized the tool and is genuinely pleased to have what feels like a colleague to assist with Latin translations."

P13, a digital humanities scholar, noted that AI's creative capabilities helped him better understand his own work, ultimately leading to a process of co-creation:

> "I was really struggling with the subtitles [for my book] ... I gave [ChatGPT] a try, and it revealed all sorts of things I didn't want, but ironically, that helped me realize what I did want to say. I also used Midjourney to generate the cover art for the book, which features an image of Queen Victoria typing on a laptop."

C-8, a data scientist, reflected on how their AI expertise shaped their relationship with AIGC technologies:

> "Before the emergence of reasoning models, I viewed AIGC as a tool. But since the release of reasoning models last November, I've started treating it as an assistant. That's why I use it every day now…I use AI to summarize literature, extract key points, accelerate reading, polish language, and translate. It's had a significant impact on my workflow. At one point, I even felt that many academic papers and books might not be necessary."

Compared with these patterns, users engaging in

**ER-4** acknowledge AI's epistemic authority in certain domains, allowing AI to lead in knowledge production. This relationship is marked by high-level trust, including appropriate and general trust, as well as process-based assessment, which often occurs in complex reasoning tasks. Users in this category often possess hybrid expertise that mixes domain knowledge and expertise in AI systems. For example, a professional programmer who is both a digital humanities domain expert and an AI expert, engages in what he described as "pair programming" with AI, treating it as a co-agent: "It's like a programming buddy… [I] accept about 70% of its suggestions. And it's completely changed the way I code" (P02-DH). The programmer also developed a sense of crisis, worrying that his job would be replaced by AI, and he had to rethink his identity as a programmer.

P12-DH also trusted AI with advanced conceptual work, using "generative AI conversationally, especially during the brainstorming stage…and it helps [her] talk through abstract ideas and clarify research questions."

While all these ERs have led to increasing depth of human-AI interaction in collaborative knowledge work, **ER-5** captures cases where users utilize AI tools but actively deny them any epistemic authority. Trust in this collaboration pattern is minimal or even absent, and the assessment remains outcome oriented. For instance, one participant described a teaching exercise in which students used generative AI to draft papers and then critique them. The AI-generated drafts typically received a "C" for lacking specificity and failing to meet rubric criteria. Despite assigning AI a complex interpretive task, the expert expressed distrust in its performance (P8-DH).

Another digital humanities scholar attributed her abstention from AI use to a deeper conviction about the purpose of research, framing her lack of trust in AI as rooted in motivational and emotional commitments to the research process:

> "I definitely think there are aspects of research that shouldn't be replaced by generative AI. To me, it comes back to a core question: why are you doing research in the first place? Using AI to streamline tedious or time-consuming tasks makes sense. But if you're outsourcing nearly every component of the process, it raises a deeper issue: do you enjoy being a researcher? If you're genuinely interested in research, the core tasks should be things you want to engage with and reflect on. So, when someone wants to replace those parts with AI, I see it as a question of purpose. For example, if I've done five interviews and need to categorize the transcripts by tomorrow, sure—using ChatGPT for efficiency makes perfect sense. But if there's no urgent deadline, why not take the time to go through the data myself? That's what makes me a better researcher. The process itself is valuable. It's why I do research in the first place." (P12-DH)

## 5. Discussion

### 5.1. Epistemic Relationship as a Situated and Negotiated Outcome

Our findings show that users' metaphoric descriptions of AI, such as a "tool" or an "assistant," do not necessarily correspond to how they actually interact with AI in specific tasks. Instead, epistemic relationships are dynamically negotiated, shifting across contexts, assessment strategies, types of trust, and user identities. Building on these insights, we propose a relational model that frames epistemic relationships not as a fixed role assignment but as a dynamic, task-contingent configuration negotiated between humans and AI systems. Our theoretical framework conceptualizes human–AI interaction as an evolving epistemic relationship shaped by users' situated engagement. Crucially, we take the epistemic capacities of AI, as discussed in recent work on AI as epistemic technology or authority (Alvarado, 2023; Ferrario et al., 2024; Hauswald, 2025), as the conceptual premise for theorizing such relationships.

In this framework, the human side of interaction is captured through observable behavioral attributes that reflect how users position and make sense of AI, including the interaction metaphor (e.g., tool, assistant, co-agent), the user's epistemic status (e.g., AI or domain expert, or layperson), the type of trust extended toward the AI (e.g., reliance, appropriate trust), and the assessment perspective (i.e., whether users evaluate AI outputs based on outcomes or underlying processes). The relationship between these dimensions is mediated by specific task contexts, which act as dynamic modulators of the interaction. Task-based variations influence the degree to which epistemic capacities are invoked or deferred, making the epistemic relationship context-sensitive and fluid rather than static.

Taken together, this model allows us to trace how AI's epistemic status is co-constructed through situated human practices, revealing not only how users metaphorically frame AI roles but also how these framings relate to underlying cognitive and epistemic commitments. Our model thus offers a more nuanced understanding of how epistemic authority, dependence, and co-agency emerge in real-world settings, emphasizing the importance of task-driven, temporally flexible, and identity-sensitive configurations in shaping meaningful human–AI knowledge partnerships. It also advances a relational epistemology that resists both technological determinism and anthropocentric essentialism, instead foregrounding the co-evolutionary nature of epistemic positioning in hybrid systems. This aligns with Hayles' (2023) critique of anthropocentrism, as our empirical cases demonstrate how epistemic authority is continuously configured through situated human–AI engagements, rather than attributed through predefined roles or human-centric criteria. By importing an epistemic perspective into HCI and operationalizing key concepts from philosophy of technology and AI epistemology, this framework also bridges abstract normative inquiry with empirical interface studies, pushing the boundaries of what it means to design and evaluate AI as a participant in sociotechnical knowledge practices.

### 5.2. Implications for the Future of Work

Our study, particularly the five epistemic relationships (ERs) we identified, captures the evolving dynamics of human–AI interaction in knowledge-intensive settings and offers broad implications for the design of workflows, training, governance, and organizational practices beyond higher education. First, the findings of our work can support more effective task allocation by aligning AI deployment with task characteristics and user needs. Each ER type corresponds to a distinct workflow model—for example, ER-1 suits structured, goal-driven tasks, while ER-3 fits exploratory, interpretive work requiring iterative human–AI interaction. These early insights provide a foundation for developing "task × epistemic relationship" matrices to guide more nuanced and context-sensitive AI integration.

Second, recognizing a user's ER type can inform more tailored AI literacy training. For example, users in ER-2 or ER-4 may benefit from greater transparency and understanding around AI reasoning and limitations to foster process-based trust. Those in ER-5 require institutional safeguards to define epistemic boundaries and prevent overreliance or misplaced responsibility.

Third, our findings also challenge the binary notion of "trust in AI." The ERs reflect a spectrum of trust configurations, each with differing accountability implications. While ER-1 and ER-2 maintain clear human responsibility, ER-5 nevertheless requires explicit mechanisms to govern trust and ensure transparent responsibility in AI-assisted decision-making.

Finally, the dynamic nature of ERs suggests job roles are epistemically fluid, not fixed. As users gain experience and engage more critically with AI, they may shift from ER-1 toward ER-3 or ER-5. This calls for organizations to support employees' evolving epistemic trajectories, enabling adaptive co-development of human roles and AI capabilities.

### 5.3 Limitations and Next Steps

As an exploratory effort, this study is not without limitations. First, our framework is inductively derived from interview data, which limits generalizability. While the five ERs offer a useful vocabulary for conceptualizing human–AI epistemic configurations, future studies are needed to validate and refine these categories across diverse domains. Second, our approach lacks behavioral or experimental measurement, which would be necessary to more precisely assess users' trust types, evaluation strategies, and epistemic expectations. Third, our findings suggest that epistemic relationships are not static but change over time, yet our current data offers only limited insight into the trajectories or transitions across ER types. Future longitudinal or process-tracing studies could illuminate the mechanisms through which users shift from instrumental reliance to co-agency, or from epistemic abstention to authority displacement. Lastly, while we hint at a relational epistemology that resists both technological determinism and anthropocentric essentialism, further theoretical work is needed to articulate how epistemic agency can be jointly constructed in hybrid systems. In particular, the conditions under which AI systems may come to be seen not only as tools, but as epistemic collaborators or authorities, remain a critical area for philosophical and empirical exploration.

## 6. Conclusion

This exploratory study brought recent philosophical discussions about AI's epistemic status, a still developing and contested topic, into the more established HCI discourse, which has traditionally focused on functionality, usability, and adoption. While many HCI studies conceptualize AI through static functional roles, our work foregrounded the epistemic dimension of AI and the dynamic ways in which users relate to AI in epistemic practices. By identifying five types of epistemic relationships, we provided a conceptual bridge between normative theories of epistemic agency and the lived realities of human–AI interaction in work settings.